\documentclass[envcountsect]{llncs}
\usepackage{times}
\title{Polynomial Synthesis of Asynchronous Automata}
\author{Nicolas {\sc Baudru} \& R\'{e}mi {\sc Morin}}
\institute
  {
    Laboratoire d'Informatique Fondamentale de Marseille\\ 
   39 rue Fr\'{e}d\'{e}ric Joliot-Curie, 
    F-13453 Marseille cedex 13, France}

\newcommand{\Loc}{\mathrm{Loc}}
\newcommand{\Tri}{\triangle}
\newcommand{\Unf}{\mathrm{Unf}}

\usepackage{newalg}

\newcommand{\MGTS}{M}\newcommand{\X}{X}
\newcommand{\G}{{\cal G}}\newcommand{\Inverse}{'}

\usepackage{amsmath}
        \makeatletter
        \def\@listI{\leftmargin\leftmargini
                    \parsep 0\p@ \@plus1\p@ \@minus\p@
                    \topsep 0\p@ \@plus1\p@ 
                     \itemsep0\p@}
         \let\@listi\@listI
         \@listi
         \makeatother
\def\bar{\overline}

\def\hat{\widehat}
\def\leq{\leqslant}
\def\geq{\geqslant}

\makeatletter
\@ifundefined{IfThenElse}
        {\newcommand{\IfThenElse}[3]{\@ifundefined{#1}{#3}{#2}}}{}
\@ifundefined{Execute}
        {\newcommand{\Execute}[1]
             {\expandafter\let\expandafter\next\csname#1\endcsname\next}}{}
\makeatother
\IfThenElse{FinDuDocumentUni}
  {\typeout{Rechargement de UNIA avorte.} }
  {}

\newcommand{\FinDuDocumentUni}{} 
\newcommand{\Inclure}[2]{}       

\usepackage{latexsym}
\usepackage{amsfonts}
\usepackage{amssymb}
\usepackage{euscript}
\usepackage{eufrak}

\def\leq{\leqslant}
\def\geq{\geqslant}

\IfThenElse{spnewtheorem}
{\spnewtheorem{Definition}{Definition}[section]{\sc}{\itshape}
\spnewtheorem{Lemma}[Definition]{Lemma}{\sc}{\itshape}
\spnewtheorem{Corollary}[Definition]{Corollary}{\sc}{\itshape}
\spnewtheorem{Proposition}[Definition]{Proposition}{\sc}{\itshape}
\spnewtheorem{Theorem}[Definition]{Theorem}{\sc}{\itshape}
\spnewtheorem{Example}[Definition]{Example}{\sc}{}
\spnewtheorem{Remark}[Definition]{Remark}{\sc}{}}
{\newtheorem{Definition}{Definition}[section]
\newtheorem{Lemma}[Definition]{Lemma}

\newtheorem{Proposition}[Definition]{Proposition}
\newtheorem{Theorem}[Definition]{Theorem}

}

\newenvironment{Proof}{\par\noindent{\bf Proof.} }
                    {\hglue 0pt plus1fill $\rule{0.5em}{0.5em}$\par}

\newcommand{\SymboleDeFinUni}{\hglue 0pt plus1fill $\rule{0.5em}{0.5em}$\par}

\newcommand{\TitreAlgoUNI}[2]{{\sc Alg.}~{#1}. #2}
\newcounter{algo}

\newcommand{\ComposeAlgoSimple}[3][t]
{\begin{figure}[#1]\hspace{\fill}{\setlength{\arraycolsep}{0mm}%
\begin{tabular}{c}      #2  \\%
\refstepcounter{algo}\TitreAlgoUNI{\thealgo}{#3}%
\end{tabular}}%
\hspace*{\fill}\end{figure}}

\newcommand{\ComposeAlgoDouble}[5][t]
{\begin{figure}[#1]\hspace{\fill}{\setlength{\arraycolsep}{0mm}%
\begin{tabular}{c|c}      #2~~~&#4  \\%
\refstepcounter{algo}\TitreAlgoUNI{\thealgo}{#3}&%
\refstepcounter{algo}\TitreAlgoUNI{\thealgo}{#5}\end{tabular}}%
\hspace*{\fill}\end{figure}}

\newcounter{CompteurListeUniEquivalence}
\renewcommand{\theCompteurListeUniEquivalence}
  {{\sf(\roman{CompteurListeUniEquivalence})}} 

\makeatletter
\def\AxUni{\@ifnextchar[ 
                {\x@AxUni}
                {\y@AxUni}}
\def\x@AxUni[#1]#2{\FormatUniAxiome{#2}_{#1}}
\def\y@AxUni#1{\FormatUniAxiome{#1}}
\makeatother
\newcommand{\FormatUniAxiome}[1]{\mbox{\sf #1}}
\newcounter{CompteurListeUniAxiome}

\newenvironment{AxiomesLocaux}[1] 
        {\begin{list}
                {$\AxUni[\theCompteurListeUniAxiome]{#1}$:}
                {\usecounter{CompteurListeUniAxiome}
                        \setlength{\labelsep}{0.5em}
                        \settowidth{\labelwidth}{$\AxUni[10]{#1}$:~}
                        \setlength{\leftmargin}{\labelwidth}
                        \addtolength{\leftmargin}{0em}}}
        {\end{list}}
        {\begin{list}
                {}
                {\usecounter{CompteurListeUniAxiome}
                        \setlength{\labelsep}{0.5em}
                        \settowidth{\labelwidth}{$\AxUni[10]{#1}$:~}
                        \setlength{\leftmargin}{\labelwidth}
                        \addtolength{\leftmargin}{1em}}
                }
        {\end{list}}
\newenvironment{AxiomesLocauxSimples} 
        {\begin{list}
                {}
                {\setlength{\labelsep}{0.5em}
                \settowidth{\labelwidth}{$\AxUni{T}$:~}
                \setlength{\leftmargin}{\labelwidth}
                \addtolength{\leftmargin}{1em}
                }}
        {\end{list}}

\newcommand{\PI}{\PrimeUni}
\newcounter{CompteurPrimeUniNombre}
\newcounter{CompteurPrimeUniStyle}
\makeatletter
\def\PrimeUni{\@ifnextchar[ 
                {\x@PrimeUni}
                {\y@PrimeUni}}
\def\x@PrimeUni[#1]#2{^{\setcounter{CompteurPrimeUniStyle}{#2}%
\setcounter{CompteurPrimeUniNombre}{#1}%
\ifnum\value{CompteurPrimeUniNombre}>0%
{\ifcase\value{CompteurPrimeUniStyle}%
\or\dagger\or\ddagger\else\circ\fi
\addtocounter{CompteurPrimeUniNombre}{-1}}
\else\fi
\ifnum\value{CompteurPrimeUniNombre}>0%
{\ifcase\value{CompteurPrimeUniStyle}%
\or\dagger\or\ddagger\else\circ\fi
\addtocounter{CompteurPrimeUniNombre}{-1}}
\else\fi
\ifnum\value{CompteurPrimeUniNombre}>0%
{\ifcase\value{CompteurPrimeUniStyle}%
\or\dagger\or\ddagger\else\circ\fi
\addtocounter{CompteurPrimeUniNombre}{-1}}
\else\fi}}
\def\y@PrimeUni{\@ifnextchar \bgroup{\z@PrimeUni}{\w@PrimeUni}}
\def\z@PrimeUni#1{\x@PrimeUni[1]{#1}}
\def\w@PrimeUni{\x@PrimeUni[1]{1}}
\makeatother

\newcommand{\NouveauSymbole}[2]
        {\expandafter\def\csname #1\endcsname{#2}}
\newcommand{\NouvelleConstruction}{\newcommand}
\newcommand{\NouvelObjet}[2]
        {\expandafter\def\csname Obj#1\endcsname{\LettreObjet{#2}}}
\newcommand{\NouvelleFonction}[2]
        {\expandafter\def\csname #1\endcsname{\FonctionUn{#2}}}

\newcommand{\LettreObjet}[1]{\EuScript{#1}}

\newcommand{\FonctionUn}[1]{\mathrm{#1}}

\newcommand{\FonctionLettreUn}[1]{\EuFrak{#1}}
\newcommand{\FonctionLettreDeux}[2]%
        {\FonctionLettreUn{#1}_{\mathrm{\lowercase{#2}}}}

\newcommand{\FoncteurUn}[1]{\EuFrak{\lowercase{#1}}}

\renewcommand{\TitreAlgoUNI}[2]{{\bf Alg.~{#1}.}~ #2}
\NouveauSymbole{Clique}{\Delta}
\NouvelleConstruction{\Application}[5]
                {{\setlength{\arraycolsep}{0mm}
                        \begin{array}[t]{rrcl}
                        #1 : & #2 &~\rightarrow~& #3\\
                             & #4 &~\mapsto~& #5
                \end{array}}}

\NouvelleConstruction{\ApplicationBis}[7]
                {{\setlength{\arraycolsep}{0mm}\begin{array}[t]{rrcl}
                        #1 : & #2 &~\rightarrow~& #3\\
                             & #4 &~\mapsto~& #5\\
                             & #6 &~\mapsto~& #7
                \end{array}}}

\NouveauSymbole{EquivEtat}{\cong}
\NouvelleConstruction{\EquivAlph}[1]{{\EquivEtat_{#1}}}
\NouveauSymbole{EquivDet}{{\EquivEtat^{d}}}
\NouvelleConstruction{\EquivAlphDet}[1]{{\EquivEtat_{#1}^{d}}}
\NouvelleConstruction{\ClasseEtat}[1]{\lfloor #1 \rfloor}
\NouvelleConstruction{\ClasseAlph}[2]{{\lfloor #1 \rfloor_{#2}}}
\NouvelleConstruction{\ClasseDet}[1]{{\lfloor #1 \rfloor^{d}}}
\NouvelleConstruction{\ClasseAlphDet}[2]{{\lfloor #1 \rfloor_{#2}^{d}}}

\NouveauSymbole{Dep}{\mbox{$\not\!\|$}}
\NouveauSymbole{Ind}{\|}

\NouvelObjet{T}{T}
\NouvelleConstruction{\T}{\ObjT}
\Inclure{UNIAbrv}{\Construction{T}}

\NouvelObjet{A}{A}
\NouvelleConstruction{\A}{\ObjA}
\NouvelObjet{S}{S}

\NouveauSymbole{EtatInitial}{\imath}
\NouveauSymbole{FctEtat}{\varsigma}

\NouvelleConstruction{\Mene}[1]{\stackrel{#1}{\longrightarrow}}
\Inclure{UNICons}{\ConstructionI{Mene}{a}}
\NouvelleConstruction{\mene}{\Mene}

\NouveauSymbole{Isomorphe}{\equiv}              
\NouveauSymbole{Unite}{\mu}
\NouveauSymbole{CoUnite}{\zeta}
\def\EnsN{{\rm I\!N}}           
\NouveauSymbole{InfEve}{\prec}
\NouveauSymbole{ALPH}{\Sigma} 
\NouveauSymbole{L}{L}
\NouveauSymbole{MotVide}{\varepsilon}
\NouveauSymbole{FctAction}{\alpha} 

\NouvelleConstruction{\Monoid}[1]{#1^{\star}}
\Inclure{UNICons}{\ConstructionI{Monoid}{A}}
\NouveauSymbole {ET}{\wedge~}
\NouveauSymbole {OU}{\vee~}
\NouveauSymbole {SSI}{\Leftrightarrow}
\NouveauSymbole{App}{\rightarrow}               
\NouvelleFonction{Id}{Id}                       
\NouveauSymbole{Equiv}{\approx}         
\NouvelleConstruction{\Famille}[3]{\left(#1_{#2}\right)_{#2\in#3}}
\Inclure{UNICons}{\ConstructionIII{Famille}{e}{i}{I}}
\NouvelleConstruction{\Quotient}[2]{#1_{/#2}}
\Inclure{UNICons}{\ConstructionII{Quotient}{X}{R}}
\NouvelleConstruction{\Classe}[2]{[#1]_{#2}}
\Inclure{UNICons}{\ConstructionII{Classe}{x}{R}}
\NouvelleConstruction{\ClotureEquiv}[1]{#1^{*}}
\Inclure{UNICons}{\ConstructionI{ClotureEquiv}{R}}
\NouvelObjet{LTL}{L}
\NouveauSymbole{L}{\ObjLTL}
\NouveauSymbole{Marquage}{\mbox{M}}
\NouveauSymbole{EnsMarquages}{\mbox{Mar}}
\NouveauSymbole{EnsMarquagesAccessibles}{\mbox{MarAcc}}
\NouvelleConstruction{\Trace}[1]{[#1]}
\Inclure{UNICons}{\ConstructionI{Trace}{u}}
\NouveauSymbole{EquivTrace}
        {\sim}
\NouveauSymbole{InfTrace}{\sqsubseteq}
\NouveauSymbole{Isomorphe}{\simeq}
\NouveauSymbole{EnsMarquagesAccessibles}{\mbox{RMar}}

\NouvelObjet{N}{N}
\NouveauSymbole{N}{\ObjN}
\NouveauSymbole{Marquage}{\mbox{M}}
\NouveauSymbole{MarquageInitial}{\mbox{M}_{\textrm{in}}}
\NouveauSymbole{MarquageFinal}{\mbox{M}_{\textrm{fin}}}
\NouveauSymbole{EnsMarquages}{\mbox{Mar}}
\NouveauSymbole{EnsMarquagesAccessibles}{\mbox{MarAcc}}

\renewcommand{\Marquage}{\ensuremath{\mathrm{M}}}

\renewcommand{\T}{\mbox{$\cal T$}}
\makeatletter
\def\Produit{\@ifnextchar* 
                {\x@Produit}
                {\y@Produit}}
\def\x@Produit*#1#2#3{\prod_{#2\in #3}#1}
\def\y@Produit#1#2#3{\prod_{#2\in #3}#1_{#2}}
\makeatother

\renewcommand{\T}{{\cal T}}
\renewcommand{\Quotient}[2]{{#1\!/\!#2}}

\renewcommand{\Famille}[3]{\left(#1_{#2}\right)_{#2\in#3}}

\renewcommand{\subsubsection}[1]{\paragraph{\textbf{#1.}}}

\NouvelObjet{B}{B}
\NouvelObjet{T}{T}
\NouvelObjet{E}{E}
\NouvelleConstruction{\E}{\ObjE}
        \Inclure{UNIAbrv}{\Construction{E}}
\NouveauSymbole{C}{C}
\NouveauSymbole{Conflit}{\sharp}
\NouveauSymbole{ConflitMin}{\sharp_{\mu}}
\NouveauSymbole{Etiq}{\xi}
\NouveauSymbole{Autorise}{\vdash}
\NouvelleFonction{Config}{Cfg}
\NouvelleFonction{Chemins}{Chemins}
\NouvelleFonction{Autorises}{Aut}

\newcommand{\NouveauFoncteur}[2]
        {\expandafter\def\csname #1a#2\endcsname{\FoncteurUn{#2}}}
\renewcommand{\FoncteurUn}[1]
        {\EuFrak{\lowercase{#1}}}
\NouveauFoncteur{LES}{PCC}
\NouveauFoncteur{PCC}{LES}

\renewcommand{\Quotient}[2]{{#1\mathord/ #2}}
\renewcommand{\EnsN}{\ensuremath{\mathbb N}}

\usepackage{endnotes}

\makeatletter
\@ifundefined{IfThenElse}
        {\newcommand{\IfThenElse}[3]{\@ifundefined{#1}{#3}{#2}}}{}
\@ifundefined{Execute}
        {\newcommand{\Execute}[1]
             {\expandafter\let\expandafter\next\csname#1\endcsname\next}}{}
\@ifundefined{Inclure}{\newcommand{\Inclure}[2]{}}{}
\@ifundefined{FinDuDocumentUni}{\newcommand{\FinDuDocumentUni}{}}{}
\makeatother

\renewcommand{\Inclure}[2]
        {\IfThenElse{Inclure#1}
        {\Execute{Inclure#1}{#2}}
        {}}

\newcommand{\UNIBTitreEntete}{}
\newcommand{\UNIBTitreEnteteVerso}{}
\newcommand{\UNIBTitreReference}{}

\newcommand{\FormatEntete}
      {\sf\footnotesize \UNIBTitreEntete \hfill \thepage}
\newcommand{\FormatEnteteVerso}
      {\sf\footnotesize \thepage  \hfill \UNIBTitreEnteteVerso}
\newcommand{\FormatEnteteReference}
      {\footnotesize\underline{\UNIBTitreReference}}

\newcommand{\InclurePUBLICATION}[1]
         {\renewcommand{\FormatEntete}
           {\footnotesize\emph{\UNIBTitreEntete \hfill \thepage}\\
                              \vspace*{1mm}\hrule}
          \renewcommand{\FormatEnteteVerso}
           {\footnotesize\emph{\thepage \hfill \UNIBTitreEnteteVerso}\\
                              \vspace*{1mm}\hrule}
           \renewcommand{\FormatEnteteReference}
              {\begin{minipage}{\textwidth} \begin{center}
                    \footnotesize \UNIBTitreReference
                  \end{center}\end{minipage}}}

\makeatletter
\def\@warning#1{\typeout{LaTeX Warning [p.\thepage, l.\the\inputlineno]: #1.}}
\makeatother

\newcount\hour
\newcount\minute
\hour=\time
\divide \hour by 60
\minute=\time
\loop \ifnum \minute > 59 \advance \minute by -60 \repeat
\def\now{%
\ifnum \hour<1 12:\else\number\hour h.\fi
\ifnum \minute<10 0\fi                 
\number\minute%
}
\makeatletter
\def\enddocument{\par\FinDuDocumentUni
\typeout{UNIB - Fin du document completee.}
\@checkend{document}\clearpage\begingroup
\if@filesw \immediate\closeout\@mainaux
\def\global\@namedef##1##2{}\def\newlabel{\@testdef r}%
\def\bibcite{\@testdef b}\@tempswafalse\makeatletter\input \jobname.aux
\if@tempswa \@warning{Label(s) may have changed. Rerun to get
cross-references right}\fi\fi\endgroup\deadcycles\z@\@@end}
\makeatother

\newcommand{\InclureLABELS}
           {\let\mylabel=\label
             \def\label##1{\mylabel{##1}\InclureLabelClair{##1}}
             \newcommand{\InclureLabelClair}[1]
              {\addtoendnotes{\protect\par{\tt[##1]}=
                  \ref{##1}\dotfill\thepage\protect\par}}%
             \let\mybibitem=\bibitem
             \newcommand{\InclureBibitemClair}[1]
              {\addtoendnotes{\par\protect\hspace*{\fill}{\tt[##1]}=\cite{##1}\protect\par}}
}

\renewcommand{\InclureLABELS}
           {\let\mylabel=\label
             \def\label##1{\mylabel{##1}\InclureLabelClair{##1}}
             \newcommand{\InclureLabelClair}[1]
              {\addtoendnotes{\protect\par{\tt[##1]}=
                  \ref{##1}\dotfill\thepage\protect\par}}%
             \let\mybibitem=\bibitem
             \def\bibitem##1{\mybibitem{##1}\InclureBibitemClair{##1}}
             \newcommand{\InclureBibitemClair}[1]
              {\addtoendnotes{\par\protect\hspace*{\fill}{\tt[##1]}=\cite{##1}\protect\par}}
}

     \Inclure{ENTETE}{Polynomial Synthesis of Asynchronous Automata}
     \Inclure{REFERENCE}{HAL}

\renewenvironment{AxiomesLocaux}[1] 
        {\begin{list}
                {$\AxUni[\theCompteurListeUniAxiome]{#1}$:}
                {\usecounter{CompteurListeUniAxiome}
                        \setlength{\labelsep}{0.5em}
                        \settowidth{\labelwidth}{$\AxUni[10]{#1}$:~}
                        \setlength{\leftmargin}{\labelwidth}
                        \addtolength{\leftmargin}{0em}}}
        {\end{list}}

\newcommand{\InsererImage}[1]{}

\usepackage{pifont} 

\usepackage{version}

\excludeversion{Commentaire}
\renewcommand{\SymboleDeFinUni}{\hglue 0pt plus1fill $\rule{0.5em}{0.5em}$\par}
\renewenvironment{Proof}{\par\noindent{\bf Proof.} }
                    {\SymboleDeFinUni}

\usepackage{graphics}

\makeatletter
\def\Homotopie{\@ifnextchar[ 
                {\x@Homotopie}
                {\y@Homotopie}}
\def\y@Homotopie{\@ifnextchar \bgroup{\z@Homotopie}{\w@Homotopie}}
\def\z@Homotopie#1{\w@Homotopie_{#1}}
\def\w@Homotopie{\EquivTrace}
\def\x@Homotopie[#1]{\@ifnextchar \bgroup
        {\xz@Homotopie{#1}}
        {\stackrel{#1}{\w@Homotopie}}}
\def\xz@Homotopie#1#2{\stackrel{#1}{\w@Homotopie}_{#2}}
\makeatother

\newcommand{\B}{\ObjB}

\begin{document}
\maketitle
\thispagestyle{plain}
\begin{abstract}
Zielonka's theorem shows that each regular set of Mazurkiewicz
traces can be implemented as a system of synchronized processes
with a distributed control structure called asynchronous automaton.
This paper gives a polynomial algorithm for the synthesis of a
non-deterministic asynchronous automaton from a regular Mazurkiewicz
trace language.
This new construction is based on an unfolding approach that improves
the complexity of Zielonka's and Pighizzini's techniques in terms of
the number of states.
\end{abstract}
\textbf{Keywords}: Concurrency theory, automata, formal languages.

\section*{Introduction}
One of the major contributions in the theory of Mazurkiewicz traces
\cite{Diekert95} characterizes regular languages by means of
asynchronous automata \cite{Zielonka87} which are devices with a
distributed control structure.
So far all known constructions of asynchronous automata from regular
trace languages are quite involved and yield an exponential explosion
of the number of states \cite{Mukund94}.
Furthermore conversions of non-deterministic asynchronous automata into
deterministic ones rely on Zielonka's time-stamping function
\cite{Klarlund94,Muscholl94} and suffer from the same state-explosion problem.
Interestingly heuristics to build small deterministic asynchronous automata
were proposed recently in \cite{Stefanescu03}.

Zielonka's theorem and related techniques are fundamental tools in concurrency
theory.
For instance they are useful to compare the expressive power of classical
models of concurrency such as Petri nets, asynchronous systems, and
concurrent automata \cite{Thiagarajan02,Morin05}.
These methods have been adapted already to the construction
of communicating finite-state machines from regular sets of message sequence
charts \cite{Mukund00}.
More recently the construction of asynchronous cellular automata \cite{Cori93}
was used to implement globally-cooperative high-level
message sequence charts \cite{Genest04}. All these constructions yield an
exponential explosion of the number of local states.

In this paper we give a \emph{polynomial} construction
of non-deterministic asynchronous automata. Our algorithm starts
from the specification of a regular trace language in the form
of a possibly non-deterministic automaton.
The latter is unfolded inductively on the alphabet into an automaton
that enjoys several structural properties
(Section~\ref{S-Unfolding algorithm}).
Next the unfolding automaton is used as the common skeleton of all
local processes (Subsection~\ref{SS-Construction}).
Our algorithm is designed specifically to ensure that the number
of local states
built is polynomial in the number of global states in the specification
(Subsection~\ref{SS-Complexity analysis}).
We show how this approach subsumes the complexity of Zielonka's and
Pighizzini's constructions (Subsection~\ref{SS-Comparisons}).

\newpage

\section{Background and main result}
In this paper we fix a finite alphabet $\ALPH$
provided with a total order $\InfTrace$.
An automaton over a subset $T\subseteq\ALPH$
is a structure $\A=(Q,\EtatInitial,T,\Mene{}, F)$
where $Q$ is a \emph{finite} set of states,
$\EtatInitial\in Q$ is an initial state,
$\Mene{}\subseteq Q\times T\times Q$ is a set of transitions,
and $F\subseteq Q$ is a subset of final states.
We write $q\Mene{a} q'$ to denote $(q,a,q')\in\Mene{}$.
Then the automaton $\A$ is called \emph{deterministic} if we have
$q\Mene{a} q' \land q\Mene{a} q'' \Rightarrow q'=q''$.
For any word $u=a_1...a_n\in\Monoid\ALPH$, we write $q\Mene{u} q'$
if there are some states $q_0,q_1,...,q_n\in Q$ such that
$q=q_0\Mene{a_1}q_1...q_{n-1}\Mene{a_n}q_n=q'$. 
A state $q\in Q$ is \emph{reachable} if $\EtatInitial\Mene{u}q$ for some
$u\in\Monoid\ALPH$.
The language $L(\A)$ accepted by some automaton $\A$
consists of all words $u\in\Monoid\ALPH$ such that $\EtatInitial\Mene{u}q$
for some $q\in F$.
A subset of words $L\subseteq\Monoid\ALPH$ is \emph{regular} if
it is accepted by some automaton.

\subsection{Mazurkiewicz traces}
We fix an \emph{independence relation} $\Ind$ over $\ALPH$,
that is, a binary relation $\Ind\subseteq\ALPH\times\ALPH$ 
which is irreflexive and symmetric.
For any subset of actions $T\subseteq\ALPH$, the \emph{dependence graph}
of $T$ is the undirected graph $(V,E)$ whose set of vertices is $V=T$
and whose edges denote dependence, i.e.\ $\{a,b\}\in E \Leftrightarrow a\Dep b$. 

The \emph{trace equivalence} $\EquivTrace$ associated
with the independence alphabet $(\ALPH,\Ind)$ is the
least congruence over $\Monoid\ALPH$ such that $ab\EquivTrace ba$
for all pairs of independent actions $a\Ind b$.
For a word $u\in\Monoid\ALPH$, the \emph{trace}
$\Trace{u}=\{v\in\Monoid\ALPH~|~ v\EquivTrace u\}$ 
collects all words that are equivalent to $u$.
We extend this notation from words to sets of words in a natural
way: For all $L\subseteq\Monoid\ALPH$,
we put $\Trace{L}=\{v\in\Monoid\ALPH ~|~\exists u\in L, v\EquivTrace u\}$.

A \emph{trace language} is a subset of words $L\subseteq\Monoid\ALPH$
that is closed for trace equivalence:
$u\in L \land v\EquivTrace u \Rightarrow v\in L$.
Equivalently we require that $L=\Trace{L}$.
With no surprise a trace language $L$ is called \emph{regular} if it is accepted
by some automaton.




\subsection{Asynchronous systems vs. asynchronous automata}
Two classical automata-based models are known to correspond to regular trace
languages. Let us first recall the basic notion of asynchronous systems
\cite{Bednarczyk88}.

\begin{Definition}\label{D-Asynchronous system}
  An automaton $\A=(Q,\EtatInitial,\ALPH,\Mene{},F)$ over the alphabet $\ALPH$
  is called an
  \emph{asynchronous system} over $(\ALPH,\Ind)$ if we have
  \begin{AxiomesLocauxSimples}
    \item[ID] $q_1\Mene{a} q_2 \land q_2\Mene{b} q_3 \land a\Ind b$
      implies $q_1\Mene{b} q_4 \land q_4 \Mene{a} q_3$ for some $q_4\in Q$.
  \end{AxiomesLocauxSimples}
\end{Definition}
The Independent Diamond property $\AxUni{ID}$ ensures that
the language $L(\A)$ of any asynchronous system is closed for the
commutation of independent adjacent actions. Thus it is a regular trace
language. Conversely it is easy to observe that \emph{any regular trace language
is the language of some deterministic asynchronous system.}

We recall now a more involved model of communicating processes known as
asynchronous automata \cite{Zielonka87}. 
A finite family $\delta=\Famille{\ALPH}{k}{K}$ of subsets of $\ALPH$
is called a \emph{distribution of $(\ALPH,\Ind)$} if we have
$a\Dep b \Leftrightarrow \exists k\in K, \{a,b\}\subseteq \ALPH_k$
for all actions $a,b\in\ALPH$.
Note that each subset $\ALPH_k$ is a clique of the dependence graph
$(\ALPH,\Dep)$ and a distribution  $\delta$
is simply a clique covering of $(\ALPH,\Dep)$.
We fix an arbitrary distribution $\delta=\Famille{\ALPH}{k}{K}$
in the rest of this paper.
We call \emph{processes} the elements of $K$.
The \emph{location} $\Loc(a)$ of an action $a\in\ALPH$ consists of
all processes $k\in K$ such that $a\in\ALPH_k$:
$\Loc(a)=\{k\in K~|~a\in\ALPH_k\}$.

\begin{Definition}
  An \emph{asynchronous automaton} over the distribution
  $\Famille{\ALPH}{k}{K}$ consists of a family of 
  finite sets of states $\Famille{Q}{k}{K}$,
  a family of initial local states $\Famille{\EtatInitial}{k}{K}$
  with $\EtatInitial_k\in Q_k$,
  a subset of final global states $F\subseteq \prod_{k\in K} Q_k$,
  and  a transition relation
  $\partial_a \subseteq\prod_{k\in \Loc(a)} Q_k\times\prod_{k\in \Loc(a)} Q_k$
  for each action $a\in\ALPH$.
\end{Definition}
The set of \emph{global states} $Q=\prod_{k\in K} Q_k$ can be provided
with a set of global transitions $\Mene{}$ in such a way that an asynchronous
automaton is viewed as a particular automaton.
Given an action $a\in \ALPH$ and two global states
$q=\Famille{q}{k}{K}$ and $r=\Famille{r}{k}{K}$, we put
$q\Mene{a}r$ if $(\Famille{q}{k}{\Loc(a)},\Famille{r}{k}{\Loc(a)})\in\partial_a$
and $q_k=r_k$ for all $k\in K\setminus\Loc(a)$.
The initial global state $\EtatInitial$ consists of
the collection of initial local states: $\EtatInitial=\Famille{\EtatInitial}{k}{K}$.
Then the \emph{global automaton} $\A=(Q,\EtatInitial,\ALPH,\Mene{},F)$
satisfies Property $\AxUni{ID}$ of
Def.~\ref{D-Asynchronous system}.
Thus it is an asynchronous system over $(\ALPH,\Ind)$
and $L(\A)$ is a regular trace language.
An asynchronous automaton is \emph{deterministic} if its global automaton
is deterministic, i.e.\ the local transition relations $\partial_a$
are partial functions.


\subsection{Main result and comparisons to related works}
\label{SS-Comparisons}
Although deterministic asynchronous automata appear as a restricted subclass of
deterministic asynchronous systems, 
Zielonka's theorem asserts that any regular trace
language can be implemented in the form of a deterministic asynchronous automaton.
\begin{Theorem}\label{T-Zielonka}\cite{Zielonka87}
  For any regular trace language $L$ there exists a
  deterministic asynchronous automaton whose global automaton $\A$
  satisfies $L=L(\A)$.
\end{Theorem}
In \cite{Mukund94} a complexity  analysis of Zielonka's construction
is detailed. Let $|Q|$ be the number of states of the minimal deterministic automaton
that accepts $L$ and $|K|$ be the number of processes. Then the number of local states
built by Zielonka's technique in each process $k\in K$ is $|Q_k|\leq 2^{O(2^{|K|}.|Q|\log |Q|)}$.
The simplified construction by Cori et al. in \cite{Cori93} also suffers from
this exponential state-explosion \cite{Diekert95}.

Another construction proposed by Pighizzini \cite{Pighizzini93} builds some non-deterministic
asynchronous automata from particular rational expressions that refine
Ochma\'nski's theorem \cite{Ochmanski85}.
This simpler approach proceeds inductively on the structure of the rational expression.
Each step can easily be shown to be polynomial. 
In particular the number of local states in each process is (at least) \emph{doubled} by each
restricted iteration. 
Consequently in some cases the number of local states in each process is
\emph{exponential} in the length of the rational expression.

In the present paper we give a new construction that
is \emph{polynomial in $|Q|$} (Th.~\ref{T-Main result}):
It produces $|Q_k|\leq O(|Q|^d)$ local states for each
process, where $d=(2.|\ALPH|+2)^{|\ALPH|+1}$,
$|\ALPH|$ is the size of $\ALPH$, and $|Q|$ is the number of states of
some (possibly non-deterministic) asynchronous system that accepts $L$.
Noteworthy the number of local states $|Q_k|$ obtained by our
approach is independent from the number of processes $|K|$.



\section{Unfolding algorithm}
\label{S-Unfolding algorithm}

In the rest of the paper we fix some automaton $\A=(Q,\EtatInitial,\ALPH,\Mene{},F)$
that is possibly non-deterministic.
The aim of this section is to associate with $\A$ a family of automata called
\emph{boxes} and \emph{triangles} which are defined inductively.
The last box built by this construction will be called the \emph{unfolding}
of $\A$ (Def.~\ref{D-Unfolding}).

Boxes and triangles are related to $\A$ by means of morphisms which are
defined as follows.
Let  $\A_1=(Q_1,\EtatInitial_1,T,\Mene{}_1,F_1)$ and
$\A_2=(Q_2,\EtatInitial_2,T,\Mene{}_2,F_2)$ be two automata over a subset of actions
$T\subseteq\ALPH$.
A \emph{morphism $\sigma:\A_1\App\A_2$ from $\A_1$ to $\A_2$} is
a mapping $\sigma:Q_1\App Q_2$ from $Q_1$ to $Q_2$
such that $\sigma(\EtatInitial_1)=\EtatInitial_2$, $\sigma(F_1)\subseteq F_2$,
and $q_1\Mene{a}_1 q_1'$ implies $\sigma(q_1)\Mene{a}_2 \sigma(q_1')$.
In particular, $L(\A_1)\subseteq L(\A_2)$.

Now boxes and triangles are associated with an initial state that
may not correspond to the initial state of $\A$.
They are associated also with a subset of actions $T\subseteq\ALPH$.
For these reasons, for any state $q\in Q$ and any subset of actions
$T\subseteq \ALPH$, we let $\A_{T,q}$ denote the automaton
$(Q,q,T,\Mene{}_T,F)$ where $\Mene{}_T$ is the restriction of $\Mene{}$
to the transitions labeled by actions in $T$: 
{ $\Mene{}_T=\Mene{}\cap (Q\times T \times Q)$.}

In this section we shall define the box $\Box_{T,q}$ for all states
$q\in Q $ and all subsets of actions $T\subseteq \ALPH$.
The box $\Box_{T,q}$ is a pair $(\B_{T,q},\beta_{T,q})$
where $\B_{T,q}$ is an automaton over $T$
and $\beta_{T,q}:\B_{T,q}\App \A_{T,q}$ is a
morphism.
Similarly, we shall define the triangle $\Tri_{T,q}$
for all states $q$ and all \emph{non-empty} subsets of transitions $T$.
The triangle $\Tri_{T,q}$ is a pair $(\T_{T,q},\tau_{T,q})$
where $\T_{T,q}$ is an automaton over $T$
and $\tau_{T,q}:\T_{T,q}\App \A_{T,q}$ is a morphism.

The \emph{height} of a box $\Box_{T,q}$ or a triangle
$\Tri_{T,q}$ is the cardinality of $T$.
Boxes and triangles are defined inductively.
We first define the box  $\Box_{\emptyset,q}$ for all
states $q\in Q$.
Then triangles of height $h$ are built upon boxes of
height $g<h$ and boxes of height $h$ are built upon either triangles
of height $h$ or boxes of height $g<h$, whether the dependence
graph $(T,\Dep)$ is connected or not.

The base case deals with the boxes of height 0.
For all states $q\in Q$, the box $\Box_{\emptyset,q}$ consists of the morphism
$\beta_{\emptyset,q}:\{q\}\App Q$ that maps $q$ to itself
together with the automaton
$\B_{\emptyset,q}=(\{q\},q,\emptyset,\emptyset,F_{\emptyset,q})$
where $F_{\emptyset,q}=\{q\}$ if $q\in F$ and $F_{\emptyset,q}=\emptyset$
otherwise.
More generally a state of a box or a triangle is final if it is associated
with a final state of $\A$.

\subsection{Building triangles from boxes}
Triangles are made of boxes of lower height.
Boxes are inserted into a triangle inductively on the height
along a tree-like structure and several copies of the same box
may appear within a triangle.
We want to keep track of this structure in order to prove
properties of triangles (and boxes) inductively.
This enables us also to allow for the distinction of different
copies of the same box within a triangle.

To do this, each state of a triangle is associated
with a \emph{rank} $k\in\EnsN$ such that all states with the same
rank come from the same copy of the same box.
It is also important to keep track of the height each state
comes from, because boxes of a triangle are included inductively
on the height.
For these reasons, a state of a triangle $\Tri_{T\PI{3},q\PI{3}}
=(\T_{T\PI{3},q\PI{3}},\tau_{T\PI{3},q\PI{3}})$ is encoded as a quadruple
$v=(w,T,q,k)$ such that $w$ is a state from the box
$\Box_{T,q}$ with height $h=|T|$ and $v$ is added to the triangle
within the $k$-th box inserted into the triangle.
Moreover this box is a copy of $\Box_{T,q}$.
In that case the state $v$ maps to $\tau_{T\PI{3},q\PI{3}}(v)=\beta_{T,q}(w)$,
that is, the insertion of boxes preserves the correspondance to the
states of $\A$.
Moreover the morphism $\tau_{T\PI{3},q\PI{3}}$ of a triangle $\Tri_{T\PI{3},q\PI{3}}$
is encoded in the data structure of its states.

\ComposeAlgoSimple
{\hspace*{\fill}\begin{minipage}{1pt}
\renewcommand{\algcommentlength}{3.5cm}
\begin{algorithm}{Build-Triangle}{T\PI{3},q\PI{3}}
  (\B,\beta)\=\CALL{Build-Box}(\emptyset, q\PI{3})\\
  (\T,\tau)\= \CALL{Mark}( (\B,\beta), \emptyset, q\PI{3}, 1)\\
  k \= 1 \\
  \begin{FOR}{h \=1 \TO |T\PI{3}|-1 \label{Al-Build-Triangle induction}}
    \begin{FOR}{v=(w,T,q,l)\text{ a state of } \T \text{ with }|T|=h-1}
      \begin{FOR}{q'\in Q  \text{and} a\in T\PI{3}\setminus T
                                                \label{Al-Build-Triangle test}}
         \begin{IF}{\beta_{T,q}(w) \Mene{a}q'}
           T' \= T\cup \{a\} \\[Here $|T'|=h<|T\PI{3}|$]
           (\B,\beta)\=\CALL{Build-Box}(T', q')
           \\[Compute $\Box_{T', q'}$]
           k \= k+1\\
           (\B',\beta')\=\CALL{Mark}((\B,\beta), T',  q', k)
           \\[Mark it with $T',q',k$]
           \CALL{Insert}((\T,\tau),(\B',\beta'))
           \\[Insert it into $(\T,\tau)$ \label{Al-Build-Triangle insert}]
           \CALL{Add}((\T,\tau),(v,a,(\EtatInitial_{\Box,T', q'},T',q',k)))
           \label{Al-Build-Triangle add}
         \end{IF}
      \end{FOR}
    \end{FOR}
  \end{FOR}\\
  \RETURN (\T,\tau)
\end{algorithm}\end{minipage}\hspace*{\fill}
\\
\centerline{\begin{minipage}{0.8\textwidth}
  N.B. Line 12, $\EtatInitial_{\Box,T', q'}$ denote the initial state of
  the box $\Box_{T', q'}$.\\
\end{minipage}}}
{Construction of a triangle \label{Al-Construction of a triangle}}

The construction of the triangle $\Tri_{T\PI{3},q\PI{3}}$
is detailed in Algorithm~\ref{Al-Construction of a triangle}.
It relies on four procedures:
\begin{itemize}
\item
  {\sc Build-Box}$(T,q)$ returns the box $\Box_{T,q}$.
\item
  {\sc Mark}$((\B,\beta),T,q,k)$ returns a copy of $(\B,\beta)$
  where each state $w$ from $\B$ is replaced by the marked state $v=(w,T,q,k)$.
\item
  {\sc Insert}$((\T,\tau),(\B,\beta))$ inserts $(\B,\beta)$
  within $(\T,\tau)$; the initial state of this disjoint union of automata
  is the initial state of $(\T,\tau)$.
\item
  {\sc Add}$((\T,\tau),(v,a,v'))$ adds a new transition $v\Mene{a} v'$
  to the automaton $\T$; it is required that $v$ and $v'$ be states of $\T$.
\end{itemize}

The construction of the triangle $\Tri_{T\PI{3},q\PI{3}}$
starts with building a copy of the base box $\Box_{\emptyset, q\PI{3}}$
which gets rank $k=1$ and whose marked initial state
$(\EtatInitial_{\Box,\emptyset,q\PI{3}},\emptyset,q\PI{3},1)$
becomes the initial
state of $\Tri_{T\PI{3},q\PI{3}}$. Along the construction of this triangle,
$k$ counts the number of boxes already inserted in the triangle.
The insertion of boxes proceeds inductively on the height $h$
(Line~\ref{Al-Build-Triangle induction}) as follows:
For each state $v=(w,T,q,l)$ with height $|T|=h-1$, if a
transition $\beta_{T,q}(w)\Mene{a}q'$ in $\A$ carries an action $a\in T\PI{3}\setminus T$
(Line~\ref{Al-Build-Triangle test}) then a new box $\Box_{T',q'}$ of height $h$ is
inserted with $T'=T\cup\{a\}$
(Line~\ref{Al-Build-Triangle insert}) and a transition $v\Mene{a} v'$ is added to the triangle
$\T_{T\PI{3},q\PI{3}}$ in construction (Line~\ref{Al-Build-Triangle add})
where $v'$ is the marked initial state of the new box $\Box_{T',q'}$.
We stress here that $\tau(v)\Mene{a}\tau(v')$ is a transition of $\A_{T\PI{3},q\PI{3}}$
because $\tau(v)=\beta_{T,q}(w)$ and
$\tau(v')=\beta_{T',q'}(\EtatInitial_{\Box,T',q'})=q'$.
This observation will show that $\tau$ is a morphism.
Another useful remark is the following.
\begin{Lemma}
  If a word $u\in\Monoid\ALPH$ leads in the triangle $\Tri_{T\PI{3},q\PI{3}}$ from its initial state
  $(\EtatInitial_{\Box,\emptyset,q\PI{3}},\emptyset,q\PI{3},1)$
  to some state  $v=(w,T,q,l)$ then each action of $\,T$ occurs in $u$.
\end{Lemma}

\ComposeAlgoSimple[b]
{\hspace*{\fill}\begin{minipage}{1pt}
\renewcommand{\algcommentlength}{4cm}
\begin{algorithm}{Missing}{T\PI{3},q,q'}
  M \= \emptyset\\
  (\T,\tau)\=\CALL{Build-Triangle}(T\PI{3}, q)\\[\label{Al-Missing aide}]
  \begin{FOR}{w\in Q_{\Tri,T\PI{3},q} \text{such that} w=(w'',T'',q'',k'')
      \text{and} |T''|=|T\PI{3}|-1}
    \begin{IF}{\tau_{T\PI{3},q}(w)\Mene{a}q'\text{with} a\in T\PI{3}\setminus T''}
          M\= M \cup\{ (w,a) \}
  \end{IF}
  \end{FOR}\\
  \RETURN (M)
\end{algorithm}\end{minipage}\hspace*{\fill}
\\
\centerline{\begin{minipage}{0.8\textwidth}
  N.B. The triangle $\Tri_{T\PI{3},q}=(\T,\tau)$ computed at
  Line~\ref{Al-Missing aide}
  consists of a set of states $Q_{\Tri,T\PI{3},q}$ and a transition
  relation $\Mene{}_{\Tri,T\PI{3},q}$.\\
\end{minipage}}}
{Set of missing transitions
from a triangle $\Tri_{T^\circ,q}$  to some state $q'$
\label{Al-Missing transitions}}

\ComposeAlgoDouble%
{\hspace*{\fill}\begin{minipage}{1pt}
\renewcommand{\algcommentlength}{0cm}
\begin{algorithm}{Min-Rank}{T\PI{3},q,\B,k}
  f\= k+1\\
  \begin{FOR}{v=(w,T',q',l) \IN \B}
    \begin{IF}{q'=q \text{ and } T'=T\PI{3}}
      \begin{IF}{l<f}
        f=l
      \end{IF}
    \end{IF}
  \end{FOR}\\
  \RETURN (f)
\end{algorithm}\end{minipage}\hspace*{\fill}}
{Minimal rank of $\Tri_{T^\circ,q}$ in $\B$~~~~\label{Al-First occurrence}}
{\hspace*{\fill}\hspace*{1cm}\begin{minipage}{1pt}
\renewcommand{\algcommentlength}{0cm}
\begin{algorithm}{Max-Out-Degree}{T\PI{3}}
  m \= 0 \\
  \begin{FOR}{q,q' \in Q}
    n \= |\CALL{Missing}(T\PI{3},q,q')|\\
    \begin{IF}{n>m}
      m\= n
    \end{IF}
  \end{FOR}\\    
  \RETURN (m)
\end{algorithm}\end{minipage}\hspace*{\fill}}
{Minimal number of copies required \label{Al-Number of copies}}

\ComposeAlgoSimple[t]
{\hspace*{\fill}\begin{minipage}{1pt}
\renewcommand{\algcommentlength}{4cm}
\begin{algorithm}{Build-Box}{T\PI{3},q\PI{3}}
  \begin{IF}{T\PI{3}=\emptyset\label{Al-Build-Box base case}}\\[This is the base case ]
    \RETURN (\CALL{Base-Box}(q\PI{3}))
  \end{IF}\\
  \begin{IF}{T\PI{3}\text{is connected (and non-empty)}\label{Al-Build-Box connected case}}\\[]    
    (\B,\beta)\= \CALL{Empty-Box}\label{Al-Build-Box connected case: initialise}
    \\[Initialise $(\B,\beta)$ to be]
    k \= 0 \\[the special empty box]
    m \= \CALL{Max-Out-Degree}(T\PI{3}) +1\\[\label{Al-Build-Box connected case: compute m}]
    \begin{FOR}{q \in Q \text{ starting with }q\PI{3}\label{Al-Build-Box connected case: begin insert}}
      (\T,\tau)\=\CALL{Build-Triangle}(T\PI{3}, q)
      \\[Compute $\Tri_{T\PI{3}, q}$]
      \begin{FOR}{l \= 1 \TO m}
        k \= k+1 \\[Insert $m$ copies]
        (\T',\tau')\=\CALL{Mark}((\T,\tau), T\PI{3},  q, k)
        \\[Marked with $T\PI{3},q,k$]
        \CALL{Insert}((\B,\beta),(\T',\tau')) \label{Al-Build-Box connected case: end insert}
      \end{FOR}
    \end{FOR}\\
    \begin{FOR}{q,q' \in Q\label{Al-Build-Box connected case: begin add}}
      M\=\CALL{Missing}(T\PI{3},q,q')\\[List of missing transitions]
      f\=\CALL{Min-Rank}(T\PI{3},q,\B,k) -1
      \\[Minimal rank of $\Tri_{T\PI{3},q}$]
      f'\=\CALL{Min-Rank}(T\PI{3},q',\B,k) -1
      \\[Minimal rank of $\Tri_{T\PI{3},q'}$]
      \begin{FOR}{j \= 1 \TO m}
        c\= 0\\[We have $|M|+1\leq m$]
        \begin{FOR}{(w,a) \in M}
          c\= c+1\\[\label{Al-Build-Box connected case: begin add bis}
          If $q=q'$ then $f=f'$]
          \begin{IF}{f+j=f'+c}
            c\=c+1
          \end{IF}\\[We have $c\leq m$]
          \CALL{Add}( (\B,\beta),
          ((w,T\PI{3},q,f+j),a,(\EtatInitial_{\Tri,T\PI{3},q'},T\PI{3},q',f'+c)))
          \label{Al-Build-Box connected case: end add}
        \end{FOR}       
      \end{FOR}       
    \end{FOR}\\
    \CALL{Clean}(\B,\beta)\\[\label{Al-Build-Box connected case: Clean}]
    \RETURN (\B,\beta)
    \end{IF}\\
    \begin{IF}{T\PI{3}\text{is not connected (nor empty)}}\\[\label{Al-Build-Box unconnected case}]    
        T_1\=\CALL{Decomposition}(T\PI{3})\\
        T_2\=T\PI{3}\setminus T_1\\
        (\B_0,\beta_0)\= \CALL{Build-Box}(T_2,q\PI{3})\\
        (\B,\beta)\=\CALL{Mark}((\B_0,\beta_0), T_2,  q\PI{3}, 1)\\
        k \= 1 \\
        \begin{FOR}{w\in Q_{\Box,T_2,q\PI{3}} \text{,} q'\in Q
            \text{ and } a\in T_1
            \label{Al-Build-Box unconnected case: Etats}}
          \begin{IF}{\beta_0(w)\Mene{a}q'}
            k \= k+1 \\[Insert a copy of $\Box_{T_1,q'}$ ]
            (\B',\beta')\= \CALL{Build-Box}(T_1,q')\\
            (\B'',\beta'')\=\CALL{Mark}((\B',\beta'), T_1,  q', k)\\
            \CALL{Insert}((\B,\beta),(\B'',\beta''))\\            
            \CALL{Add}( (\B,\beta),
            ((w,T_2,q\PI{3},1),a,(\EtatInitial_{\Box,T_1,q'},T_1,q',k)))
            \label{Al-Build-Box unconnected case: Etat initial}
          \end{IF}
        \end{FOR}\\
      \RETURN (\B,\beta)        
  \end{IF}
\end{algorithm}\end{minipage}\hspace*{\fill}
\\
\centerline{\begin{minipage}{0.8\textwidth}
  \begin{itemize}
  \item[N.B.]
  \item In Line~\ref{Al-Build-Box connected case: end add}
    $\EtatInitial_{\Tri,T\PI{3},q'}$ 
    denotes the initial state of $\Tri_{T\PI{3},q'}$.
  \item In Line~\ref{Al-Build-Box unconnected case: Etats}
    $Q_{\Box,T_2,q\PI{3}}$ denotes the set of states of $\Box_{T_2,q\PI{3}}$.  
  \item
    In Line~\ref{Al-Build-Box unconnected case: Etat initial}
    $\EtatInitial_{\Box,T_1,q'}$ denotes the initial state of $\Box_{T_1,q'}$.
  \end{itemize}~
\end{minipage}}}
{Construction of a box \label{Al-Construction of a box}}


\subsection{Building boxes from triangles}
We distinguish two cases when we build the box $\Box_{T,q}$
whether the dependence graph $(T,\Dep)$ is connected or not.
In case  $(T,\Dep)$ is a connected graph then  the box $\Box_{T,q}$
collects all triangles $\Tri_{T,q'}$ for all states $q'\in Q$.
Each triangle is duplicated a fixed number of times
and copies of triangles are connected in some particular way.
Similarly to triangles, the states of a box are decorated
with a rank $k$ that distinguishes states from different
triangles and also states from different copies of the same triangle.
We adopt the same data structure as for triangles:
A state $v$ of a box is a quadruple
$(w,T,q,k)$ where $w$ is a state of $\Tri_{T,q}$ and $k\in\EnsN$.
Whereas triangles of height $h$ are built upon boxes of height $g<h$,
boxes $\Box_{T,q}$ are built upon triangles $\Tri_{T,q'}$
with the same set of transitions $T$ ---~and consequently,
with the same height.
Similarly to the algorithm {\sc Build-Triangle}, the algorithm
that builds boxes uses an integer variable $k$ that counts the
number of triangles already inserted in the box in construction.

In case the dependence graph $(T,\Dep)$ is not connected,
we let $T_1$ denote the connected component of $(T,\Dep)$ that contains the least
action $a\in T$ w.r.t.\ the total order $\InfTrace$ over $\ALPH$
and we put $T_2= T\setminus T_1$.
Then  the box $\Box_{T,q}$ is built upon a copy of the box
$\Box_{T_2,q}$ connected to copies of boxes $\Box_{T_1,q_1}$
for some states $q_1\in Q$.

The construction of the box $\Box_{T\PI{3},q\PI{3}}$
is detailed in Algorithm~\ref{Al-Construction of a box}.
It relies on ten procedures:
\begin{itemize}
\item
  {\sc Base-Box}$(q)$ returns the base box $\Box_{\emptyset,q}$.
\item
  {\sc Empty-Box} returns a special new box called \emph{empty box}.
\item
  {\sc Mark}, {\sc Insert} and {\sc Add} are the procedures used
  for {\sc Build-Triangle}.
  If $(\B,\beta)$ is this special empty box then 
  {\sc Insert}$((\B,\beta),(\T,\tau))$ replaces simply $(\B,\beta)$
  by $(\T,\tau)$.
\item
  {\sc Missing}$(T\PI{3},q,q')$ returns the set of all pairs
  $(w,a)$ where $w$ is a state 
  that has been inserted in the triangle $\Tri_{T\PI{3},q}$
  within a box $\Box_{T'',q''}$ such that $|T''|=|T\PI{3}|-1$
  and the action $a\in T\PI{3}\setminus T''$ is such that there is a transition
  $\tau_{T\PI{3},q}(w)\Mene{a}q'$ in $\A$ (Alg.~\ref{Al-Missing transitions}).
  Due to the structure of triangles, if $(w,a)$ is a missing transition
  then there is no transition
  $w\Mene{a}_{\Tri,T\PI{3},q}w'$ with $\tau_{T\PI{3},q}(w')=q'$
  in $\Tri_{T\PI{3},q}$.
\item
  {\sc Min-Rank}$(T\PI{3},q,\B,k)$ returns the minimal rank of
  a copy of a triangle $\T_{T\PI{3},q}$ inserted in $\B$ where $k$ is the maximal
  rank of triangles in $\B$
  (Alg.~\ref{Al-First occurrence}).
\item
  {\sc Max-Out-Degree}$(T\PI{3})$ returns the number of copies of each
  triangle $\Tri_{T\PI{3},q}$ that should compose the box
  $\Box_{T\PI{3},q\PI{3}}$. It does not depend on $q$ but it
  depends on the cardinality of all sets
  {\sc Missing}$(T\PI{3},q,q')$ with $q,q'\in Q$
  (Alg.~\ref{Al-Number of copies}).
  The r\^{o}le of these copies is detailed below.
\item
  {\sc Clean}$(\B,\beta)$ remove all unreachable states from $\B$.
\item
  {\sc Decomposition}$(T\PI{3})$ returns
  the connected component $T$ of $(T\PI{3},\Dep)$ that contains the minimal
  action of $T\PI{3}$ w.r.t.\ the total order $\InfTrace$.
\end{itemize}

The construction of the box $\Box_{T\PI{3},q\PI{3}}$
starts with solving the base case where $T\PI{3}=\emptyset$
(Line~\ref{Al-Build-Box base case}).
Assume now that the dependence graph $(T\PI{3},\Dep)$ is connected
(Line~\ref{Al-Build-Box connected case}).
Then the box is initialized as the special empty box
(Line~\ref{Al-Build-Box connected case: initialise}).
The number $m$ of copies of each triangle $\Tri_{T\PI{3},q}$
is computed in Line~\ref{Al-Build-Box connected case: compute m}
with the help of functions {\sc Max-Out-Degree} and {\sc Missing}.
Next these copies are inserted and the
first copy of $\Tri_{T\PI{3},q\PI{3}}$ gets rank $k=1$
(Lines~\ref{Al-Build-Box connected case: begin insert} to
\ref{Al-Build-Box connected case: end insert}). 
Consequently the initial state of the box $\Box_{T\PI{3},q\PI{3}}$
in construction is the first copy of the initial state $
\EtatInitial_{\Tri, T\PI{3},q\PI{3}}$ of the triangle
$\Tri_{T\PI{3},q\PI{3}}$, that is:
$(\EtatInitial_{\Tri, T\PI{3},q\PI{3}}, T\PI{3},q\PI{3}, 1)$.
Noteworthy copies of the same triangle have
consecutive ranks.

In a second step transitions are added to connect these triangles
to each other (Lines \ref{Al-Build-Box connected case: begin add} to
\ref{Al-Build-Box connected case: end add}). 
Intuitively a $a$-transition is \emph{missing} from the state $w=(w'',T'',q'',k'')$
of the triangle $\Tri_{T\PI{3},q}$ to the state $q'$ of $\A$ if
$|T\PI{3}\setminus T''|=1$ ---~i.e.\ this state has been inserted at the highest level
in $\Tri_{T\PI{3},q}$~---
and there exists in $\A$ a transition $\tau_{T\PI{3},q}(w)\Mene{a} q'$ with
$a\in T\PI{3}\setminus T''$ but no transition
$w\Mene{a} w'$ with $\tau_{T\PI{3},q}(w')=q'$ in $\Tri_{T\PI{3},q}$.

The r\^{o}le of {\sc Missing} is to compute the missing transitions w.r.t.\ $q$, $q'$,
and $T\PI{3}$.
For each such missing transition $(w,a)$ we connect each copy of $w$ to the initial
state of a copy of $\Tri_{T\PI{3},q'}$.
In this process we require two crucial properties:
\begin{AxiomesLocaux}{P}
\item
  No added transition connects two states from the same copy of the
  same triangle: $(w,T\PI{3},q,l)$ should not be connected to
  $(\EtatInitial_{\Tri, T\PI{3},q}, T\PI{3},q, l)$.
\item
  At most one transition connects one copy of $\Tri_{T\PI{3},q}$
  to one copy of $\Tri_{T\PI{3},q'}$:
  If we add from a given copy of $\Tri_{T\PI{3},q}$ a transition
  $(w_1,T\PI{3},q,l)\Mene{a}(\EtatInitial_{\Tri, T\PI{3},q'},T\PI{3},q',l')$
  and a transition
  $(w_2,T\PI{3},q,l)\Mene{b}(\EtatInitial_{\Tri, T\PI{3},q'},T\PI{3},q',l')$
  to the same copy of $\Tri_{T\PI{3},q'}$ then $w_1=w_2$ and $a=b$.
\end{AxiomesLocaux}

The minimal number of copies required to fulfill these conditions is computed
by {\sc Max-Out-Degree}.
For a fixed missing transition $(w,a)$ from a state $w$
of the triangle $\Tri_{T\PI{3},q}$ to a state $q'$ of $\A$,
Lines~\ref{Al-Build-Box connected case: begin add bis} to
\ref{Al-Build-Box connected case: end add} add a
transition from the $j$-th copy of $w$ to the
$c$-th copy of the initial state of $\Tri_{T\PI{3},q'}$
with the property that $j\not=c$ if $q=q'$ (Condition~$\AxUni[1]{P}$ above).
Moreover states from the $j$-th copy of $\Tri_{T\PI{3},q}$ are connected to
distinct copies of the initial state of $\Tri_{T\PI{3},q'}$
(Condition~$\AxUni[2]{P}$ above).

Note here that each new transition $(v,a,v')$ added to
$(\B,\beta)$ at Line~\ref{Al-Build-Box connected case: end add}
is such that
$\beta(v)\Mene{a}\beta(v')$ is a transition from  $\A_{T\PI{3},q\PI{3}}$
because
$\beta(v)=\tau_{T\PI{3},q}(w)$,
$\beta(v')=\tau_{T\PI{3},q'}(\EtatInitial_{\Tri,T\PI{3},q'})=q'$,
and $\tau_{T\PI{3},q}(w)\Mene{a}q'$.
Again, this observation will show that $\beta$ is a morphism.
A crucial remark for boxes of connected alphabets is the following.
\begin{Lemma}
  If a non-empty word $u$ leads from the initial state of a triangle
  $\Tri_{T\PI{3},q}$ to the initial state of a triangle $\Tri_{T\PI{3},q'}$
  within the box $\Box_{T\PI{3},q\PI{3}}$ then each action of $T\PI{3}$ occurs
  in $u$.
\end{Lemma}

For simplicity's sake our algorithm uses the same number of copies for each
triangle. This approach yields in general unreachable states in useless copies.
The latter are removed by {\sc Clean} at Line~\ref{Al-Build-Box connected case: Clean}.

Assume now that $(T\PI{3},\Dep)$ is not connected
(Line~\ref{Al-Build-Box unconnected case}).
Let $T_1$ be the connected component of $T\PI{3}$ that contains the
least action of $T\PI{3}$ w.r.t.\ the total order $\InfTrace$ over $\ALPH$.
We put $T_2=T\PI{3}\setminus T_1$.
The construction of the box $\Box_{T\PI{3},q\PI{3}}$ starts with building
a copy of the box $\Box_{T_2,q\PI{3}}$.
Next for each state $w$ of $\Box_{T_2,q\PI{3}}$ and each transition
$\beta_{T_2,q}(w)\Mene{a} q'$ with $a\in T_1$, the algorithm inserts a (new)
copy of the box $\Box_{T_1,q'}$ and adds a transition from the copy of $w$ to the
initial state of the copy of $\Box_{T_1,q'}$.
By recursive calls of {\sc Build-Box} the box $\Box_{T\PI{3},q}$ is built along a tree-like
structure upon copies of boxes $\Box_{T',q'}$ where $T'$ is a connected component of $T\PI{3}$.

\subsection{Remarks}
%
From a mathematical viewpoint, Algorithms~\ref{Al-Construction of a triangle}
to \ref{Al-Construction of a box}
are meant to define boxes $\Box_{T,q}$ and triangles $\Tri_{T,q}$.
Thus two instances of {\sc Build-Triangle}$(T,q)$
produce the same object. For this reason, we speak of \emph{the}
triangle $\Tri_{T,q}$.
This is particularly important to understand the interaction between
{\sc Build-Box} and {\sc Missing}.
In case $T$ is connected,
Algorithm {\sc Build-Box} proceeds in two steps.
First several copies of each triangle $\Tri_{T,q}$ are collected and next
some transitions are added from some states of copies of $\Tri_{T,q}$ to the
initial state of copies of $\Tri_{T,q'}$. These additional transitions are
computed in a separate function {\sc Missing} that depends on
triangles. 
It is crucial that the triangles $\Tri_{T,q}$ used by the function
{\sc Missing} be the same as the triangles $\Tri_{T,q}$ inserted in
{\sc Build-Box}.

From a more computational viewpoint, Algorithms~\ref{Al-Construction of a triangle} to
\ref{Al-Construction of a box} can obviously be implemented.
To do this, we require that each triangle and each
box be constructed only once. An alternative to this requirement is to
adapt the parameters of the function {\sc Missing}
and ensure that {\sc Build-Box} transfers its own triangle $\Tri_{T,q}$
instead of the pair $(T,q)$ to that function so that the set of states
computed by {\sc Missing} matches the set of states used by {\sc Build-Box}.
However it need not to transfert its own triangle $\Tri_{T,q}$ to the function
{\sc Max-Out-Degree} because this function works on triangles up to isomorphisms.

In this section we have built a family of boxes and triangles from a fixed
automaton $\A$. This construction leads us to the definition of the unfolding
of $\A$ as follows.
\begin{Definition}\label{D-Unfolding}
  The \emph{unfolding} $\A_\Unf$ of the automaton $\A=(Q,\EtatInitial,\ALPH,\Mene{},F)$
  is the box $\B_{\ALPH,\EtatInitial}$; moreover
  $\beta_{\Unf}$ denote the mapping $\beta_{\ALPH,\EtatInitial}$
  from the states of $\A_\Unf$ to $Q$.
\end{Definition}
In the next section we study some complexity, structural, and semantical properties
of this object.
We assume that $\A$ satisfies Property $\AxUni{ID}$ of
Definition~\ref{D-Asynchronous system} so that it accepts a regular trace language $L$. 
We explain how to build from the unfolding $\A_\Unf$ a non-deterministic
asynchronous automaton 
that accepts $L(\A)$.

\clearpage
\newpage
\section{Properties of the unfolding algorithm}
\label{S-Properties}
In this section we fix a regular trace language $L$ over the independence
alphabet $(\ALPH,\Ind)$.
We assume that the possibly non-deterministic automaton $\A$
fulfills  Property $\AxUni{ID}$ of Def.~\ref{D-Asynchronous system} and
satisfies $L(\A)=L$.
First we sketch a complexity analysis of the number of states
in the unfolding $\A_\Unf$.
Next we show in Subsection~\ref{SS-Construction}
how to build from $\A_\Unf$ an asynchronous automaton whose global automaton
accepts $L(\A)$.

\subsection{Complexity analysis}\label{SS-Complexity analysis}
For all naturals $n\geq 0$ we denote by $\beta_n$ the maximal number
of states in a box $\B_{T,q}$ with $|T|=n$ and $q\in Q$.
Similarly for all naturals $n\geq 1$ we denote by $\tau_n$ the maximal number
of states in a triangle $\T_{T,q}$ with $|T|=n$ and $q\in Q$.
Noteworthy $\beta_0=1$ and $\tau_1=1$.
Moreover $\tau_n$ is non-decreasing because the triangle $\Tri_{T',q}$ is a
subautomaton of the triangle $\Tri_{T,q}$ as soon as $T'\subseteq T$.
In the following we assume $2\leq n\leq|\ALPH|$.

Consider some subset $T\subseteq\ALPH$ with $|T|=n$.
Each triangle $\T_{T,q}$ is built inductively upon boxes
of height $h\leq n-1$ (see Alg.~\ref{Al-Construction of a triangle}).
We distinguish two cases.
First boxes of height $h<n-1$ are inserted. Each of  these boxes
appears also in some triangle $\T_{T',q}$ with $T'\subset T$ and
$|T'|=n-1$.
Each of these triangles is a subautomaton of $\T_{T,q}$ with at most
$\tau_{n-1}$ states.
Moreover there are only $n$ such triangles which give rise to
at most $n.\tau_{n-1}$ states built along this first step.
Second, boxes of height $n-1$ are inserted and connected
to states inserted at height $n-2$.
Each of these states belongs to some box $\Box_{T',q'}$ with $|T'|=n-2$;
it gives rise to at most $2.|Q|$ boxes at height $n-1$
because $|T\setminus T'|=2$: This yields at most
$2.|Q|.\beta_{n-1}$ new states. Altogether we get
\begin{equation}
  \tau_n\leq n.\tau_{n-1}.(1+2.|Q|.\beta_{n-1})\leq
  |\ALPH|.\tau_{n-1}.3.|Q|.\beta_{n-1}
\end{equation}

Consider now a connected subset $T\subseteq\ALPH$ with $|T|=n-1$.
Then each box $\B_{T,q}$ is built upon triangles $\T_{T,q'}$ of height $n-1$
(see Alg.~\ref{Al-Construction of a box}).
We can check that the value $m=${\sc Max-Out-Degree}$(T)$ is at most $\tau_{n-1}+1$.
Therefore the box $\B_{T,q}$ contains at most $2.\tau_{n-1}$ copies
of each triangle $\T_{T,q'}$. Hence
$\beta_{n-1}\leq 2.|Q|.\tau_{n-1}^2$.

Consider now a non-connected subset $T\subseteq\ALPH$ with $|T|=n-1$.
Then each box $\B_{T,q}$ is built upon copies of boxes $\B_{T',q'}$
where $T'$ is a connected component of $(T,\Dep)$.
These boxes are inserted inductively along recursive calls of
{\sc Build-Box} and they are connected in a tree-like structure.
Each of these boxes contains at most $2.|Q|.\tau_{n-2}^2$ states
as explained above. From each state of these boxes at most
$(n-2).|Q|$ new boxes are connected.
Thus each box $\B_{T',q'}$ is connected to at most
$c=|\ALPH|.2.|Q|^2.\tau_{n-2}^2$ boxes in the tree-like structure.
Consequently there are at most $1+c+c^2+c^3+...+c^{n-2}$ boxes.
It follows that
\begin{equation}
  \beta_{n-1}\leq c^{n-1}.2.|Q|.\tau_{n-2}^2
  \leq 2^n.|\ALPH|^{n-1}.|Q|^{2.n-1}.\tau_{n-2}^{2.n}
\end{equation}
Since $\tau_{n-2}\leq\tau_{n-1}$ we get in both cases
$\beta_{n-1}\leq 2^{|\ALPH|}.|\ALPH|^{|\ALPH|-1}.|Q|^{2.|\ALPH|-1}.(\tau_{n-1})^{2.|\ALPH|}$.

We can now apply $(1)$ and get
$\tau_n\leq N.\tau_{n-1}^d$ where $N=3.2^{|\ALPH|}.|\ALPH|^{|\ALPH|}.|Q|^{2.|\ALPH|}$
and $d=2.|\ALPH|+1$. Since $\tau_1=1$, we get $\tau_n\leq N^{d^{n-1}}$.
We can apply $(2)$ with $n$ instead of $n-1$ and get
$\beta_n\leq 2.|Q|.N.(\tau_n)^{2.(n+1)}\leq  2.|Q|.N.N^{d^{n-1}.(2n+2)}$.
Finally we have
\begin{equation}
  \beta_{|\ALPH|}\leq 2.|Q|.\left( 3.2^{|\ALPH|}.|\ALPH|^{|\ALPH|}.|Q|^{2.|\ALPH|}
  \right)^{(2.|\ALPH|+2)^{|\ALPH|}}\in O\left(|Q|^{(2.|\ALPH|+2)^{|\ALPH|+1}}\right)
\end{equation}

\subsection{Construction of an asynchronous automaton}
\label{SS-Construction}
Finally we build from the unfolding
$\A_\Unf=(Q_{\Unf},\EtatInitial_{\Unf},\ALPH,\Mene{}_{\Unf},F_{\Unf})$
of $\A$ an asynchronous automaton $\hat{\A_\Unf}$ that accepts $L(\A)$.
We define $\hat{\A_\Unf}$ as follows.
First we put $Q_k=Q_{\Unf}$ for each process $k\in K$.
Next the initial state is the $|K|$-tuple
$(\EtatInitial_{\Unf},...,\EtatInitial_{\Unf})$.
Moreover for each action $a$, the pair
$(\Famille{q}{k}{\Loc(a)},\Famille{r}{k}{\Loc(a)})$
belongs to the transition relation $\partial_a$
if there exist two states $q,r\in Q_{\Unf}$
and a transition $q\Mene{a}_{\Unf} r$ in the unfolding
such that the two following conditions are satisfied:
\begin{itemize}
\item for all $k\in\Loc(a)$, $q_k\Mene{u}_{\Unf} q$ for some word
  $u\in\Monoid{(\ALPH\setminus\ALPH_k)}$;
\item for all $k\in\Loc(a)$, $r_k=r$; in particular all $r_k$ are equal.
\end{itemize}
Finally, a global state $\Famille{q}{k}{K}$ is \emph{final} if
there exists a final state $q\in Q_{\Unf}$ such that
for all $k\in K$ there exists a path  $q_k\Mene{u}_{\Unf} q$
for some word $u\in\Monoid{(\ALPH\setminus\ALPH_k)}$.

\begin{Theorem}\label{T-Main result}
  The asynchronous automaton $\hat{\A_\Unf}$ satisfies
  $L(\hat{\A_\Unf})=L(\A)$. Moreover the number of local states $Q_k$
  in each process is polynomial in $|Q|$ where $|Q|$ is the number of states
  in $\A$; more precisely $|Q_k|\leq O(|Q|^d)$ where $d=(2.|\ALPH|+2)^{|\ALPH|+1}$.
\end{Theorem}

\subsection{Sketch of proof}
By induction on the structure of the unfolding it is not difficult
to check the following first property (see Appendix~\ref{S-Appendix semantical}).
\begin{Lemma}\label{L-Semantics property}
   The mapping $\beta_{\Unf}$ is a morphism from the unfolding $\A_\Unf$ to $\A$.
   Moreover for all $u\in L(\A)$ there exists $v\in L(\A_\Unf)$ such that
   $v\EquivTrace u$.
\end{Lemma}

The proof of Theorem~\ref{T-Main result} relies on an intermediate asynchronous
automaton $\bar{\A_\Unf}$ over some extended independence alphabet
$(\bar\ALPH,\bar\Ind)$.
We consider the alphabet
$\bar\ALPH = \ALPH \cup \{ (a,k)\in \ALPH\times K~|~a\not\in \ALPH_k \}$
provided with the independence relation $\bar{\Ind}$ such that
$a\bar\Dep b$ iff $a\Dep b$,
$a\bar\Dep (b,k)$ iff $a\in \ALPH_k$, and
$(a,k)\bar\Dep (b,k')$ iff $k=k'$
for all actions $a,b\in\ALPH$ and all processes $k,k'\in K$.
For each process $k\in K$ we put
$\bar{\ALPH_k}=\ALPH_k\cup \{(a,k)~|~a\in\ALPH\setminus\ALPH_k \}$.
It is easy to check that $\Famille{\bar\ALPH}{k}{K}$ is a distribution of
$(\bar\ALPH,\bar\Ind)$.
Now $\bar{\A_\Unf}$ shares with $\hat{\A_\Unf}$ its local states $Q_k$ and its initial state.
A global state $\Famille{q}{k}{K}$ is final if there exists a final state $q\in F_\Unf$
of the unfolding such that $q_k=q$ for all $k\in K$.
 For each action $a\in\ALPH$ its transition relation
$\bar\partial_a$ is such that
$((q_k)_{k\in\Loc(a)},(q'_k)_{k\in\Loc(a)})\in \bar\partial_a$
if there exists a transition $q\Mene{a}_\Unf q'$ such that
$q_k=q$ and $q'_k=q'$ for all $k\in\Loc(a)$.
Moreover for each internal action $(a,k)\in\bar\ALPH\setminus\ALPH$,
we put $(q,q')\in\bar\partial_{(a,k)}$ if $q\Mene{a}_\Unf q'$.

We consider the projection morphism
$\rho:\Monoid{\bar\ALPH}\App\Monoid\ALPH$ such that
$\rho(\MotVide)=\MotVide$, $\rho(u.a)=\rho(u).a$ if $a\in\ALPH$, and
$\rho(u.a)=\rho(u)$ if $a\in\bar\ALPH\setminus\ALPH$.
It is not difficult to prove that $L(\A_\Unf)\subseteq\rho(L(\bar{\A_\Unf}))$
and $\rho(L(\bar{\A_\Unf}))=L(\hat{\A_\Unf})$. These two basic properties
do not rely on the particular structure of $\A_\Unf$:
They hold actually for any automaton. 

On the contrary the proof of the next lemma is very technical and
tedious and relies on the particular construction of boxes and triangles
(see Appendix~\ref{S-Appendix main} for some details).
\begin{Lemma}\label{L-Key lemma}
  For each box $\Box_{T,q}=(\B_{T,q},\beta_{T,q})$
  we have $\rho(L(\bar{\B_{T,q}}))\subseteq\Trace{L(\B_{T,q})}$.
\end{Lemma}
We can now conclude.
By Lemma~\ref{L-Semantics property} we have
$\Trace{L(\A_\Unf)}=L(\A)$.
On the other hand the two basic properties show that
$\Trace{L(\A_\Unf)}\subseteq L(\hat{\A_\Unf})$.
Conversely Lemma~\ref{L-Key lemma} yields
$L(\hat{\A_\Unf})=\rho(L(\bar{\A_\Unf}))\subseteq \Trace{L(\A_\Unf)}$.
Therefore $L(\A)=L(\hat{\A_\Unf})$.

\newpage

\section*{Conclusion and future work}
We have presented a polynomial algorithm for the construction of
non-determi\-nistic asynchronous automata from regular trace languages.
We have shown that this new unfolding method improves the complexity
of known techniques in terms of the number of local and global states.
Several variations of our approach lead to analogous complexity results.
We have selected here the simplest version to analyse. But it might not
be the more efficient in practice.

Interestingly this unfolding method can be adapted to the implementation of
any globally-cooperative compositional
high-level message sequence charts as investigated in \cite{Genest04}.
At present we are developping more involved unfolding techniques
in order to construct deterministic safe asynchronous automata
\cite{Stefanescu03}.
We are also investigating a possible extension of our unfolding technique
to infinite traces \cite{Gastin92}.

\newpage
\appendix

\section{Proof of Lemma~\ref{L-Semantics property}}
\label{S-Appendix semantical}

An immediate induction shows that for each box $\Box_{T,q}=(\B_{T,q},\beta_{T,q})$
and each triangle $\Tri_{T,q}=(\T_{T,q},\tau_{T,q})$, the mappings
$\beta_{T,q}$ and $\tau_{T,q}$ are morphisms from
$\B_{T,q}$ to $\A_{T,q}$ and from $\T_{T,q}$ to $\A_{T,q}$
respectively.
In particular $L(\A_\Unf)\subseteq L(\A)$.

By induction on the size of $T$ we prove that
for all paths $q\Mene{u} q_1$ of $\A_{T,q}$ there exists an equivalent 
word $u'\EquivTrace u$ such that 
$\EtatInitial_{\Box,T,q}\Mene{u'} v$ is a path of $\B_{T,q}$ and $\beta_{T,q}(v) = q_1$.
This property is trivial for the empty set $T=\emptyset$
because $\A_{\emptyset,q}$ and  $\B_{\emptyset,q}$ are reduced to  the state $q$.
We shall distinguish two cases whether $T$ is connected or not.

Assume first that $T$ is a connected set of actions.
We proceed by induction on the length of $u$.
The property holds for the empty word because $\beta_{T,q}(\EtatInitial_{\Box,T,q}) = q$.
Let $q \Mene{u} q_1\Mene{a}q_2$ be a  path of $\A_{T,q}$.
By induction there is $u'\EquivTrace u$ such that 
$\EtatInitial_{\Box,T,q}\Mene{u'} v$ is a path in $\B_{T,q}$
and $\beta_{T,q}(v) = q_1$. 
Then, by construction, 
the state $v = (w,T,q',k)$ comes from some triangle $\Tri_{T,q'}$.
Furthermore $w$ comes from a box $\Box_{T'',q''}$ inserted in
$\Tri_{T,q'}$: We have
$w=(w'',T'',q'',k'')$.
We distinguish several cases.
\begin{enumerate}
\item 
  If $a\in T\setminus T''$ and $|T\setminus T''|=1$. Then
  $(w,a)$ belongs to $\mbox{\sc{Missing}}(T,q',q_2)$.
  Consequently
  Line~$25$ of Alg.~\ref{Al-Construction of a box} shows that
  $v \Mene{a} (\EtatInitial_{\Tri,T,q_2},T,q_2,k')$ for some integer $k'$ and 
  $\beta_{T,q} (\EtatInitial_{\Tri,T,q_2},T,q_2,k') = \tau_{T,q_2}(\EtatInitial_{\Tri,T,q_2}) = q_2$.
\item
        If  $a\in T\setminus T''$ and $|T\setminus T''|\ne 1$.
        Then Line~$13$ of Alg.~\ref{Al-Construction of a triangle} shows that
        $w\Mene{a} w'$ with $w' = (\EtatInitial_{\Box,T''\cup \{a\},q_2},T''\cup \{a\},q_2,k')$ 
        for some integer $k'$   is a transition of $\T_{T,q'}$ and $\tau_{T,q'}(w') = q_2$.
        Consequently $v\Mene{a} (w',T,q',k)$ is a transition of $\B_{T,q}$
        and $\beta_{T,q}(w',T,q',k) = q_2$ (see Line~$14$ of Alg.~\ref{Al-Construction of a box}).
\item
        If  $a \in T''$. By construction the path $\EtatInitial_{\Box,T,q}\Mene{u'} v$ 
        of $\B_{T,q}$ consists of the sequence of transitions 
        $\EtatInitial_{\Box,T,q}\Mene{u_1}v_1\Mene{u_2} v$ 
        such that $u_1.u_2 = u'$,
        $v_1 = (w_1,T,q',k)$, $w_1 = (\EtatInitial_{\Box,T'',q''},T'',q'',k'')$ 
        and all states $v_2$ reach along the path $v_1\Mene{u_2} v$ 
        come from the same  box $\Box_{T'',q''}$ 
        of the same triangle $\Tri_{T,q'}$ that is 
        $v_2$ is some tuple  $((w_2,T'',q'',k''),T,q',k)$. 
        Consequently each action $b$ that occurs in $u_2$ belongs to $T''$:
        It follows that $q''\Mene{u_2}q_1\Mene{a}q_2$ is a path of $\A_{T'',q''}$.
        By induction there is an equivalent
        word $u'_2\sim u_2.a$ such that $\EtatInitial_{\Box,T'',q''}\Mene{u'_2}w'$
        is a path of $\B_{T'',q''}$ and $\beta_{T'',q''}(w') = q_2$.
        Consequently  
        $v_1\Mene{u'_2}((w',T'',q'',k''),T,q',k)$
        is a path of $\B_{T,q}$ and 
        $\beta_{T,q}((w',T'',q'',k''),T,q',k) = q_2$
        (see Line~$12$ of Alg.~\ref{Al-Construction of a triangle} 
        and Line~$14$ of Alg.~\ref{Al-Construction of a box}).
\end{enumerate}

Suppose now that $T$ is  an unconnected set of actions. 
Let $q\Mene{u} q_1$ be a  path of $\A_{T,q}$,
$T_1$ be the connected component that contains the least action of $T$
and $T_2 = T\setminus T_1$.
If $u|T_1 = \varepsilon$ then $q \Mene{u} q_1$
is also a path of $\A_{T_2,q}$. Consequently, by induction 
 there exists $u'\EquivTrace u$ such that 
$\EtatInitial_{\Box,T_2,q}\Mene{u'} w$ is a path of $\B_{T_2,q}$ and $\beta_{T_2,q}(w) = q_1$.
It follows by Line~$33$ of Alg.~\ref{Al-Construction of a box} that 
$\EtatInitial_{\Box,T,q}\Mene{u'} (w,T_2,q,1)$ is a path of $\B_{T,q}$
 and $\beta_{T,q}(w,T_2,q,1) = q_1$.
If $u|T_1  = a.u_1$ and $u|T_2  = u_2$
then $q\Mene{u_2} q_2 \Mene{a} q_3 \Mene{u_1} q_1$ 
is also a path of $\A_{T,q}$ because $u_2.a.u_1 \sim u$ and
$\A$ satisfies $\AxUni{ID}$.
Moreover $q\Mene{u_2} q_2$ is  a path of $\A_{T_2,q}$
and $q_3 \Mene{u_1} q_1$ is a path of $\A_{T_1,q_3}$.
Consequently, by induction, 
 there exists $u'_2\EquivTrace u_2$ such that 
$\EtatInitial_{\Box,T_2,q}\Mene{u_2'} w_2$ is a path of $\B_{T_2,q}$ and $\beta_{T_2,q}(w_2) = q_2$,
and on other hand, 
 there exists $u'_1\EquivTrace u_1$ such that 
$\EtatInitial_{\Box,T_1,q_3}\Mene{u'_1} w_1$ is a path of $\B_{T_1,q_3}$ 
and $\beta_{T_1,q_3}(w_1) = q_1$.
Then Alg.~\ref{Al-Construction of a box} ensures that 
$\EtatInitial_{\Box,T,q}\Mene{u'_2} (w_2,T_2,q,1)$ is a path of $\B_{T,q}$, 
$\beta_{T,q}(w_2,T_2,q,1) = q_2$ (Line~$33$),
$(\EtatInitial_{\Box,T_1,q_3},T_1,q_3,k)\Mene{u'_1} (w_1,T_1,q_3,k)$ is a path of $\B_{T,q}$
for some integer $k$ and $\beta_{T,q}(w_1,T_1,q_3,k) = q_1$ (Line~$40$).
Finally we have
$(w_2,T_2,q,1) \Mene{a} (\EtatInitial_{\Box,T_1,q_3},T_1,q_3,k)$ (Line~$35$ and $41$).
It follows that $\EtatInitial_{\Box,T,q}\Mene{u'_2.a.u'_1} (w_1,T_1,q_3,k)$
is a path of $\B_{T,q}$, $\beta_{T,q}(w_1,T_1,q_3,k) = q_1$ and $u'_2.a.u'_1 \sim u$.


\section{Proof sketch of Lemma~\ref{L-Key lemma}}
\label{S-Appendix main}
The complete proof of Lemma~\ref{L-Key lemma} requires about 20 pages
of tedious technical details. In this appendix we present the two main
ideas that lead the argument.
We need first to introduce some basic definitions and notations
precisely.

\paragraph{Some basic definitions and notations.}
Let $\A$ be some automaton over $\ALPH$.
A \emph{path} of length $n\in\EnsN\setminus\{0\}$
is a sequence of transitions $\Famille{q_i\Mene{a_i} q'}{i}{[1,n]}$
such that $q'_i=q_{i+1}$ for all integers $0<i<n$.
For all words $u\in\Monoid\ALPH$ we write $q\Mene{u} q'$ to denote
a path $\Famille{q_i\Mene{a_i} q'}{i}{[1,n]}$ where $q_1=q$, $q'_n=q'$,
and $u=a_1...a_n$.
Then $q$ is called the domain of $q\Mene{u} q'$ and $q'$ is called its
codomain.
A path of length 0 is simply a state $q$ of $\A$. Its domain and codomain
are equal to $q$.

If $s$ and $s'$ are two paths such that the codomain of $s$ is the domain
of $s'$ then the \emph{product} $s\cdot s'$ is defined in a natural way:
If the  length of $s$ is 0 then $s\cdot s'=s'$;
if the  length of $s'$ is 0 then  $s\cdot s'=s$;
otherwise $s\cdot s'$ is the concatenation of $s$ and $s'$.

Note that if $s$ is a path of the length $l>0$ then it is the product
of two paths $s=s_1\cdot s_2$ where the length of $s_1$ is $1$
and the length of $s_2$ is $l-1$.
Moreover such a product is unique.
This remark allows us to define mappings for paths inductively on the length.

\paragraph{Projections of global states and executions.}
Assume now that $\A$ is (the global system of) an asynchronous automaton
over the distribution $\Famille{\ALPH}{k}{K}$.
Then a path of $\A$ is called an \emph{execution}.
For convenience we shall consider the component automata $\Famille{\A}{k}{K}$
defined as follows: For each process $j\in K$,
$\A_j=(Q_j,\EtatInitial_j,\ALPH_j,\Mene{}_j,Q_j)$
where $q_j\Mene{a}_j q'_j$ if there are
$q=\Famille{q}{k}{\Loc(a)}$ and $q=\Famille{q'}{k}{\Loc(a)}$ such that
$(q,q')\in\partial_a$. Note here that $j\in\Loc(a)$ since $a\in\ALPH_j$.

Now the \emph{projection $s|k$} of an execution $s$ of $\A$
onto a process $j\in K$ is a path of $\A_j$ defined inductively as follows: 
\begin{itemize}
\item $s|j = q_j$ if $s$ is a path of length 0 that corresponds to the
  global state $(q_k)_{k\in K}$;
\item $s|j = q_j\Mene{a} q'_j\cdot(s'|j)$ if $s$ is the product
  $s=t\cdot s'$ where $t$ is a transition $(q_k)_{k\in K} \Mene{a} (q'_k)_{k\in K}$
  and $j\in \Loc(a)$.
\item $s|j = s'|j$ if $s$ is the product $s=t\cdot s'$ where 
  $t$ is a transition $q\Mene{a} q'$ and $j\not\in \Loc(a)$.
\end{itemize}

\paragraph{Executions of extended asynchronous automata.}
\newcommand{\barbox}[1]{\stackrel{\bar{~~}}{#1}}
\newcommand{\ELoc}{\mathbb{L}}

In the paper we define the \emph{extended asynchronous automaton} $\bar{\A_\Unf}$
of the unfolding automaton $\A_\Unf$. This definition can naturally be generalized
to any automaton.
Let $\A$ be an automaton and $\bar \A$ the corresponding extended asynchronous
automaton.
We say that an execution $s =  q\Mene{u}  q'$ of $\bar \A$ is \emph{arched}
if there are two states $v$ and $v'$ in $\A$ such that 
for all $k\in K$, $ q|k = v$ and $ q'|k = v'$.
Noteworthy each execution that leads an extended asynchronous automaton 
$\bar \A$ from its global initial state to some global final state is arched.

We define now a function $\gamma$ that associates each action of $\bar\ALPH$
to the corresponding action of $\ALPH$ in a natural way:
For all actions $(a,k)\in \bar\ALPH\setminus \ALPH$, $\gamma(a,k) = a$ 
and for all actions $a\in \ALPH$, $\gamma(a) = a$.
As usual this map extends from actions to words and we get
$\gamma:\Monoid{\bar\ALPH}\App\Monoid{\ALPH}$.
We can also extend the mapping $\gamma$ as a function from paths of
component automata $\A_k$ to paths of $\A$ as follows.
For each sequence $s$ that is a path of some $\A_k$, we define
$\gamma(s)$ inductively on the length of $s$ by
\begin{itemize}
\item $\gamma(s)=q$ if the length of $s$ is 0 and $s=q$.
\item $\gamma(s)=q\Mene{\gamma(a)} q'\cdot \gamma(s')$ if $s$ is a product
$s=t\cdot s'$ where $t$ is a transition $q\Mene{a} q'$.
\end{itemize}
Clearly if $s$ is an execution of $\bar\A$ and $k$ a process of $K$ then 
$s|k$ is a path of $\A_k$ and $\gamma(s|k)$ is a path of $\A$.

\paragraph{Definitions associated to unfoldings.}
Let $T$ be a non-empty subset of $\ALPH$.
We consider the triangle
$\T_{T,q} = (Q_{\Tri,T,q},\EtatInitial_{\Tri,T,q},\Mene{}_{\Tri,T,q},F_{\Tri,T,q})$.
Let $v$ be a state from $\T_{T,q}$. 
By construction of $\T_{T,q}$, $v$ is a quadruple $(w,T',q',k')$ 
such that $w$ is a state from the box $\Box_{T',q'}$ and $k'\in\EnsN$. 
We say that the \emph{box location} of $v$ is $l^\Box(v) = (T',q',k')$.
We define the \emph{sequence of boxes} reached along a path $s=q\Mene{u}q'$
in  $\T_{T,q}$ as follows:
\begin{itemize}
\item If the length of $s$ is $0$ and $s$ corresponds to state $q\in Q_{\Tri,T,q}$
  then $\ELoc^\Box(s)=l^\Box(q)$.
\item If $s$ is a product $s=s'\cdot t$ where
  $t$ is the transition $q\Mene{a}q'$ then
  two cases appear:
  \begin{itemize}
  \item If $l^\Box(q)=l^\Box(q')$ then $\ELoc^\Box(s)=\ELoc^\Box(s')$;
  \item If $l^\Box(q)\not=l^\Box(q')$ then $\ELoc^\Box(s)=\ELoc^\Box(s').l^\Box(q')$
  \end{itemize}
\end{itemize}

Similarly we define  the sequence of triangles $\ELoc^\Tri(s_1)$ reached
by a path $s_1$ in a box $\B_{T_1,q_1}$ where $T_1$ is a non-empty \emph{connected}
set of actions and the sequence of boxes $\ELoc^\Box(s_2)$ reached by a path
$s_2$ in a box $\B_{T_2,q_2}$ where $T_2$ is an \emph{unconnected}
set of actions.
 
\paragraph{Two main properties of unfoldings.}
The following proposition states that all processes behave similarly in 
an extended asynchronous automaton built from boxes or triangles.

\begin{Proposition}\label{all processes go through the same locations}
  Let $\B_{T_1,q_1}$ be a box with $T_1$ a non-empty \emph{connected} set of actions,
  $\B_{T_2,q_2}$ be a box with $T_2$ an \emph{unconnected} set of actions, and 
  $\T_{T_3,q_3}$ be a triangle with $T_3$ a non-empty set of actions.
  Let $s_1$, $s_2$ and $s_3$ be  \emph{arched} executions 
  of $\bar \B_{T_1,q_1}$,  $\bar \B_{T_2,q_2}$ and  $\bar \T_{T_3,q_3}$
  respectively. Then:
  \begin{enumerate}
  \item $\forall k,k' \in \Loc(T_1)$,
    $\ELoc^\Tri(\gamma( s_1|k)) = \ELoc^\Tri(\gamma( s_1|k'))$;
  \item $\forall k,k' \in K$,
    $\ELoc^\Box(\gamma( s_2|k)) = \ELoc^\Box(\gamma( s_2|k'))$;
  \item $\forall k,k' \in K$,
    $\ELoc^\Box(\gamma( s_3|k)) = \ELoc^\Box(\gamma( s_3|k'))$.
\end{enumerate}
\end{Proposition}
\begin{Proof}  
Property~2 and Property~3 stem from the remark that 
$\B_{T_2,q_2}$ and $\T_{T_3,q_3}$
are made of boxes connected along a tree-like structure.
The proof of Property~1 is more subtle.
Let $a$ be an action of $T_1$ and $k,k'$ be two processes of $\Loc(a)$.
We proceed by contradiction.
Let $\T$ and $\T'$ be the first triangles that differ in 
$\ELoc^\Tri(\gamma( s_1|k))$ and $\ELoc^\Tri(\gamma( s_1|k'))$.
Let $c$ be the number of $a$-transitions that occur in $ s_1$ just before 
$\gamma( s_1|k)$ and $\gamma( s_1|k')$ reach $\T$ and $\T'$.
Since $ s_1$ is arched, $\gamma( s_1|k)$ and $\gamma( s_1|k')$
have to meet eventually for the last state.
Therefore $\gamma( s_1|k)$ and $\gamma( s_1|k')$
have to leave triangles $\T$ and $\T'$  respectively. 
Consequently, there is a $(c+1)^{th}$ $a$-transition $ q\Mene{a}  q'$ 
in $ s_1$. Moreover, this transition is such that $ q |k$ is a state 
from $\T$ whereas $ q |k'$ is a state from $\T'$, 
that is:  $ q |k \ne  q |k'$.
This contradicts the definition of $\bar \partial_a$.
\end{Proof}

\begin{Proposition}\label{L-main part for box}
        Let $\B_{T,q}$ be a box.
        Let $s =  q \Mene{u}  q'$ be an arched execution of $\bar \B_{T,q}$
        with $ q |k = w$ and $ q' |k = w'$ for all $k\in K$.
        Then there is a word $v\in {\ALPH}^\star$ such that
        $v\EquivTrace \rho(u)$ and $w\Mene{v}w'$ is a path of $\B_{T,q}$.
\end{Proposition}
\begin{Proof}
        We proceed by induction on the size of $T$.
        The case where $T=\emptyset$ is trivial because 
        $\B_{\emptyset,q}$ consists of a single state $q$.
        Suppose that the property holds for all subsets $T'\subset T$
        and  all states $q'\in Q$.
        Assume first that $T$ is a connected set of actions.
        By Proposition~\ref{all processes go through the same locations},
        we know that
        $\ELoc^\Tri(\gamma(s|k)) = \ELoc^\Tri(\gamma( s|k'))$
        for all processes $k,k' \in \Loc(T_1)$.
        We claim first that we can find an other execution $s' =  q \Mene{u'}  q'$
        such that for \emph{all} processes $k,k'\in K$, 
        $\ELoc^\Tri(\gamma(s'|k)) = \ELoc^\Tri(\gamma(s'|k'))$ and
        moreover $\rho(u) = \rho(u')$.
        Let $\ELoc^\Tri(\gamma(s'|k)) = \T_1...\T_n$ be the sequence of triangles
        visited by $\gamma(s'|k)$. We can split the execution $s'$ into 
        several smaller arched executions $s_1,\dots,s_n$ such that 
        each execution $s_i$ is located within triangle $\T_i$.
        Similarly each execution $s_i$ can be split into 
        several smaller arched executions $s'_1,...,s'_m$ such that 
        each execution $s'_j$ is located within a box $\B_j$ inserted
        in $\T_i$.
        Then we can conclude by applying the inductive hypothesis
        on each smaller box.
        
        Assume finally that $T$ is an unconnected set of actions.
        By Proposition~\ref{all processes go through the same locations},
        we know that for all processes $k,k'\in K$, 
        $\ELoc^\Box(\gamma(s|k)) = \ELoc^\Box(\gamma(s|k'))$. 
        Then we can conclude by applying the inductive hypothesis
        on the smaller boxes visited by $s$.
\end{Proof}     

Lemma~\ref{L-Key lemma} follows now immediately: We have
        $\rho(L(\bar \B_{T,q})) \subseteq [ L(\B_{T,q} ) ]$.

\end{document}